\documentclass[fleqn,11pt]{article}

\usepackage{graphicx}
\usepackage{hyperref}

\usepackage{amsfonts,amssymb,latexsym,cite}

\def\noi{\noindent}

\newcommand{\Title}[1]{\noi {{\Large\bf #1}}\\[1ex]}

\newcommand{\Author}[2]{\noi{\bf #1}\\[2ex]\noi{\normalsize\it #2}\\}

\newcommand{\Abstract}[1]{\vskip 2mm \begin{center}
        \parbox{16.4cm}{\small\noi #1} \end{center}\medskip}

\newcommand{\foom}[1]{\protect\footnotemark[#1]}

\def\email#1#2{\footnotetext[#1]{e-mail: #2}\addtocounter{footnote}{1}}

\begin{document}

\twocolumn[

\Title {On collisions with unlimited energies in the vicinity of Kerr
and Schwarzschild black hole horizons}

\Author{A. A. Grib\foom 1, Yu. V. Pavlov\foom 2 and O. F. Piattella\foom 3}
       {$^{1,2}$A.Friedmann Laboratory for Theoretical Physics,
Griboedov kanal 30/32, St.\,Petersburg 191023, Russia\\
$^{1}$Theoretical Physics and Astronomy Department, The Herzen  University,
48, Moika, St.\,Petersburg, 191186, Russia\\
$^{2}$Institute of Mechanical Engineering, Russian Acad. Sci.,
Bol'shoy pr. 61, St. Petersburg 199178, Russia\\
$^{3}$Dep. de F\'isica, Universidade Federal do Esp\'irito Santo,
av. Ferrari 514, 29075-910 Vit\'oria, ES, Brazil\\
$^{3}$INFN sezione di Milano, Via Celoria 16, 20133 Milano, Italy}

\Abstract
 {Two  particle collisions close to the horizon of the rotating
nonextremal Kerr's and Schwarzschild black holes are analyzed.
    For the case of multiple collisions it is shown  that high energy in
the centre of mass frame occurs due to a great relative velocity of
two particles and a large Lorentz factor.
    The dependence of the relative velocity on the distance to horizon is
analyzed, the time of movement from the point in the accretion disc to
the point of scattering with large energy
as well as the time of back movement to the Earth are calculated.
    It is shown that they have reasonable order.}

]   
\email 1 {andrei\_grib@mail.ru}
\email 2 {yuri.pavlov@mail.ru}
\email 3 {oliver.piattella@gmail.com}

\section{Introduction}

    There is much interest today to the high energy processes in the
ergosphere of the Kerr's rotating black hole as the model for
Active Galactic Nuclei (AGN).
    In Ref.~\cite{GribPavlov2007AGN} some of the authors of this paper put
the hypothesis that due to Penrose process and scattering in the vicinity
of the horizon superheavy particles of dark matter due to the large centre
of mass energy transfer  can become ordinary particles observed on the Earth
as ultra high energy cosmic rays (UHECR) by
the AUGER group~\cite{Auger07}.

    In Ref.~\cite{BanadosSilkWest09} a resonance for the centre of mass (CM)
energy of two scattering particles close to the horizon of the extremal Kerr's
black hole was found.
    Let us call this effect the BSW effect.
    In our
papers~\cite{GribPavlov2010}--\cite{GribPavlov2011b}
it was shown that the BSW effect can occur for the nonextremal black hole if
one takes into account the possibility of multiple scattering of the particle:
in the first scattering close to the horizon the particle gets the angular
momentum close to the critical one.
    In the second scattering close to the first one the particles due to
BSW effect occur to be in the region of high energy physics --- Grand
Unification or even Planckean physics.
    In Ref.~\cite{Zaslavskii10c}
(see also Refs.~\cite{Zaslavskii10,Zaslavskii11})
it was shown that the BSW effect can be
connected with the special behaviour of the Killing vector on the ergosphere
and the large Lorentz factor for relative velocity of two particles.
    In this paper we continue our analysis of this process
made in Refs.~\cite{GribPavlov2010}--\cite{GribPavlov2011b}.

    The system of units $ G=c=1$ is used in the paper.

\section{The scattering energy in the centre of mass frame}

    The Kerr's metric of the rotating black hole in Boyer--Lindquist
coordinates has the form
    \begin{eqnarray}
\hspace*{-9pt}
d s^2 \!= d t^2 \!-
\frac{2 M r}{\rho^2} ( d t \!- a \sin^2 \! \theta\, d \varphi )^2
\nonumber \\
-\, \rho^2  \Bigl( \frac{d r^2}{\Delta} + d \theta^2 \Bigr)
- (r^2 + a^2) \sin^2 \! \theta\, d \varphi^2,
\label{Kerr}
\end{eqnarray}
    where
    \begin{equation} \label{Delta}
\rho^2 = r^2 + a^2 \cos^2 \! \theta, \ \ \ \
\Delta = r^2 - 2 M r + a^2,
\end{equation}
    $M$ is the mass of the black hole, $aM$ is angular momentum.
    The event horizon for the Kerr's black hole corresponds to the value
    \begin{equation}
r = r_H \equiv M + \sqrt{M^2 - a^2} .
\label{Hor}
\end{equation}
    The Cauchy horizon is
    \begin{equation}
r = r_C \equiv M - \sqrt{M^2 - a^2} . \label{HorCau}
\end{equation}

    For geodesics in Kerr's metric~(\ref{Kerr}) one
obtains (see Ref.~\cite{NovikovFrolov}, Sec.~3.4.1)
    \begin{equation} \label{geodKerr1}
\rho^2 \frac{d t}{d \lambda } = -a \left( a E \sin^2 \! \theta - J \right)
+ \frac{r^2 + a^2}{\Delta}\, P,
\end{equation}
    \begin{equation}
\rho^2 \frac{d \varphi}{d \lambda } =
- \left( a E - \frac{J}{\sin^2 \! \theta} \right) + \frac{a P}{\Delta} ,
\label{geodKerr2}
\end{equation}
    \begin{equation} \label{geodKerr3}
\rho^4 \left( \frac{d r}{d \lambda} \right)^2 = R, \ \ \ \
\rho^4 \left( \frac{d \theta}{d \lambda} \right)^2 = \Theta.
\end{equation}
    Here
$m$ is the rest mass of the probe particle, $\lambda = \tau /m $,
where $\tau$ is the proper time for massive particle,
$E$ is conserved energy of the probe particle,
$J$ is conserved angular momentum projection on the rotation axis
of the black hole,
    \begin{equation} \label{geodP}
P = \left( r^2 + a^2 \right) E - a J,
\end{equation}
    \begin{equation} \label{geodR}
R = P^2 - \Delta [ m^2 r^2 + (J- a E)^2 + Q],
\end{equation}
    \begin{equation} \label{geodTh}
\Theta = Q - \cos^2 \! \theta \left[ a^2 ( m^2 - E^2) +
\frac{J^2}{\sin^2 \! \theta} \right],
\end{equation}
    $Q$ is the Carter's constant.
    For massless particle one must take the limit $m \to 0$
in formulas~(\ref{geodR}), (\ref{geodTh}).
    For equatorial ($\theta=\pi/2$) geodesics the Carter constant
is equal zero and $\Theta=0$ also.

    Let us find the energy $E_{\rm c.m.}$ in the centre of mass system
of two colliding particles with rest masses~$m_1$ and~$m_2$
in arbitrary gravitational field.
    It can be obtained from
    \begin{equation} \label{SCM}
\left( E_{\rm c.m.}, 0\,,0\,,0\, \right) = p^{\,i}_{(1)} + p^{\,i}_{(2)},
\end{equation}
    where $p^{\,i}_{(n)}$ is 4-momentum of particle with number~$n$.
    Taking the squared~(\ref{SCM}) and due to $p^{\,i}_{(n)} p_{(n)i}= m_n^2$
one obtains
    \begin{equation} \label{SCM2af}
E_{\rm c.m.}^{\,2} = m_1^2 + m_2^2 + 2 p^{\,i}_{(1)} p_{(2)i} .
\end{equation}
    The scalar product does not depend on the choice of the coordinate frame
so~(\ref{SCM2af}) is valid in an arbitrary coordinate system and for arbitrary
gravitational field.

    Note that the transformation to the centre of mass coordinate system
is always possible excluding the case of two massless particles with
identically directional momenta.
    But in this case particles cannot collide.

    Let us find the expression for the collision energy of two particles
freely falling at the equatorial plane of the rotating black hole.
    We denote~$x=r/M$, \ $ A=a/M$, \ $ j_n = J_n/M$,
    \begin{equation} \label{DenKBHxhc}
x_H = 1 + \sqrt{1 - A^2}, \ \ \
x_C = 1 - \sqrt{1 - A^2},
\end{equation}
    \begin{equation} \label{Delxxhxc}
\Delta_x = x^2 - 2 x + A^2 = (x - x_H) (x- x_C) .
\end{equation}

    Using~(\ref{geodKerr1})--(\ref{geodKerr3}) one obtains:
    \begin{eqnarray}
p_{(1)}^{\,i} p_{(2) i} =
\frac{1}{x \Delta_x} \Biggl\{
E_1 E_2 \left( x^3 + A^2(x+2) \right) -
\nonumber \\
-\, 2A \left( j_1 E_2 + j_2 E_1 \right) + j_1 j_2 (2-x)-
\nonumber \\
-\, \biggl[ \biggl( 2 E_1^2 x^2 + 2 (j_1 - E_1  A )^2 - j_1^2 x
\nonumber  \\
+\, (E_1^2 - m_1^2 ) x \Delta_x
\biggr)
\biggl( 2 E_2^2 x^2 + 2 (j_2 - E_2  A )^2
\nonumber  \\
-\, j_2^2 x +
(E_2^2 - m_2^2 ) x \Delta_x \biggr) \biggr]^{\textstyle \frac{1}{2}} \Biggr\}.
\label{KerrL1L2}
\end{eqnarray}

    The value of $\Delta_x$ is going to zero on the event horizon and
as it is seen from~(\ref{KerrL1L2}) the scalar product of four
vectors $u_{(1)}^i u_{(2) i}$ and the collision energy of particles on
the horizon can be divergent depending on the behavior of the denominator
of the formula.
    To find the limit $r \to r_H$ for the black hole with a given angular
momentum~$A$ one must take in~(\ref{KerrL1L2}) $x = x_H + \alpha$
with $\alpha \to 0 $ and do calculations up to the order~$\alpha^2$.
    Taking into account $ A^2 = x_H x_C$, $x_H + x_C=2$, after resolution
of uncertainties in the limit $\alpha \to 0 $ one obtains
    \begin{eqnarray}
E_{\rm c.m.}^{\,2}(r \to r_H) = \frac{ (J_{1H} J_2 - J_{2H} J_1)^2}
{4 M^2 (J_{1H} - J_1) (J_{2H} - J_2)}
\nonumber  \\
+\, m_1^2
\left[1+ \frac{J_{2H} \!- J_2}{J_{1H} \!- J_1}\right] +
m_2^2
\left[ 1+ \frac{J_{1H} \!- J_1}{J_{2H} \!- J_2}\right]\!,\,
\label{KerrLime1e2e2f}
\end{eqnarray}
    where
    \begin{equation} \label{KerrlH} 
J_{nH} = \frac{2 E_n r_H}{A} = \frac{E_n}{\Omega_H}\,,
\end{equation}
$\Omega_H = A/2 r_H$ is horizon angular velocity~\cite{NovikovFrolov}.
    Formula (\ref{KerrLime1e2e2f}) can be used when one or both particles
become massless.
    The BSW effect can  be considered also in this case.
    In~\cite{GribPavlov2011b,HaradaKimura11} the expression of $E_{\rm c.m.}$
is written in other forms.

    For the Schwarzschild black hole $(A=0)$ the energy of collision in
the centre of mass frame is
    \begin{eqnarray}
E_{\rm c.m.}^{\,2}(r \to r_H) =
\frac{ (E_1 J_2 - E_2 J_1)^2}{4 M^2 E_1 E_2}
\nonumber  \\
+\, m_1^2 \left(1 + \frac{E_2}{E_1} \right) +
m_2^2 \left(1 + \frac{E_1}{E_2} \right).
\label{SchwLime1e2e}
\end{eqnarray}

    As it can be seen from~(\ref{KerrLime1e2e2f}) the collision energy of
particle in the centre of mass frame goes to infinity on the horizon if
the angular momentum of one of the freely falling particles has
the value~$J_{nH}$.
    Is  falling of the particle with such a value of the angular momentum
on the horizon possible?
    For the case of the free fall from infinity on the nonextremal $A<1$
rotating black hole it is impossible.
    This can be seen from the fact that the expression~(\ref{geodR})
on the horizon is going to zero for $J \to J_H$ but it's derivative
with respect to~$r$ for $A<1$ is negative.

    Consider the case of the collision of two massive particles.
    For massive particles $p^{\,i}_{(n)} = m_{(n)} u_{(n)}^{\,i} $,
where $u^i=dx^i/d \tau $.
    In this case, as one can see from~(\ref{SCM2af}),
the energy $ E_{\rm c.m.} $ has maximal value
for given $u_{(1)}, u_{(2)}$ and $ m_1 + m_2 $,
if the particle masses are equal: $m_1 =m_2$.

    Let us find the  expression of the energy in the centre of mass frame
through the relative velocity~$ v_{\rm rel}$ of particles at the moment
of collision~\cite{BanadosHassanainSilkWest10}.
    In the reference frame of the first particle one has for
the components of 4-velocities  of particles at this moment
    \begin{equation} \label{Relsk01}
u_{(1)}^i = (1,0,0,0), \ \ \
u_{(2)}^i = \frac{(1, \, \mathbf{v}_{\rm rel}) }{\sqrt{1\!- v_{\rm rel}^2}}.
\end{equation}
    So \ $ u_{(1)}^i u_{(2) i} = 1 \left/ \sqrt{1- v_{\rm rel}^2}\right. $,
    \begin{equation} \label{Relsk02}
v_{\rm rel} = \sqrt{1- \left( u_{(1)}^i u_{(2) i} \right)^{-2}}\,.
\end{equation}
    These expressions evidently don't depend on the coordinate system.

    From~(\ref{SCM2af}) and~(\ref{Relsk02}) one obtains
    \begin{equation} \label{Relsk03}
E_{\rm c.m.}^{\,2} = m_1^2 + m_2^2 +
\frac{2 m_1 m_2}{\sqrt{1 \!- v_{\rm rel}^2}}
\end{equation}
    and the nonlimited growth of the collision energy in the centre of mass
frame occurs due to growth of the relative velocity to the velocity of
light~\cite{Zaslavskii11}.

    The massive particle free falling in the black hole with dimensionless
angular momentum~$A$ being nonrelativistic at infinity $(E = m) $
to achieve the horizon of the black hole must have angular momentum
from the interval $(l_L,\, l_R)$,
    \begin{equation} \label{geodKerr5}
l_L = - 2 [ 1 + \sqrt{1+A}\, ], \ \
l_R = 2 [ 1 + \sqrt{1-A}\, ]
\end{equation}
(here the notation $l= J/mM$ is used).
    For $A<1$ one has $l_R < l_H$ and
even for values close to the extremal $A=1$ of the rotating black hole
$E_{\rm c.m.}^{\, \rm max}/ \sqrt{ m_1 m_2} $ can be not very large
as mentioned in Refs.~\cite{BertiCardosoGPS09}, \cite{JacobsonSotiriou10}
for the case $m_1=m_2$.
    So for $A_{\rm max} =0.998 $ considered as the maximal possible
dimensionless angular momentum of the astrophysical black holes
(see Ref.~\cite{Thorne74}) one obtains
$ E_{\rm c.m.}^{\, \rm max} /\sqrt{m_1 m_2}\approx 18.97 $.
    However this evaluation is enough for collisions of superheavy particles
of dark matter with the mass close to the Grand Unification scale to occur
in the region of Grand Unification interaction physics so that these
particles can decay on quarks and be observed as
the UHECR~\cite{GribPavlov2007AGN}.

    Does it mean that in processes of usual particles (protons, electrons)
scattering in the vicinity of the rotating nonextremal black holes
the scattering energy is limited so that no Grand Unification or
even Planckean energies can be obtained?
    As it was shown for the first time in~\cite{GribPavlov2010}
if one takes into account the possibility of multiple scattering so that
the particle falling from infinity on the black hole with some fixed
angular momentum changes its momentum in the result of interaction with
particles in the accreting disc and after this is again scattering close to
the horizon then the scattering energy can be unlimited.

    Note that the critical values of $J_{H}$ have various values for
different values of the energies $E$ (see~(\ref{KerrlH})).
    That is why there is not only one critical trajectory
for collision with unlimited energy  but a set of trajectories corresponding
to an interval of specific energies of particles falling onto a black hole.

    Let us get the expression for the permitted interval in~$r$ for
particles with angular momentum $l = l_H - \delta $ close to the horizon.
    From~(\ref{geodKerr3}), (\ref{geodP}), (\ref{geodR}) one has for
the movement of massive particle in the equatorial plane
    \begin{equation} \label{geodKerr3dop}
\hspace*{-9pt}
\left( \frac{d r}{d \tau} \right)^2 = \varepsilon^2 +
\frac{2}{x^3} \, (A \varepsilon - l)^2 +
\frac{A^2 \varepsilon^2 \!- l^2}{x^2} - \frac{\Delta_x}{x^2},
\end{equation}
    where $\varepsilon =E/m$ is specific energy of particle.
    As it was shown in Ref.~\cite{LL_II}, Sec.~88,
the specific energy in the static gravitational field is  equal to
    \begin{equation} \label{bh1}
\varepsilon = \sqrt{ \frac{ g_{00} }{1 -\mathbf{v}^{2} } } ,
\end{equation}
    where $\mathbf v$ is the velocity of the particle measured by
the observer at rest at the point of the passing particle.

    To get the boundaries of the permitted interval in $x$
one must put the left hand side of~(\ref{geodKerr3dop}) to zero and
find the root.
    In the second order in~$\delta$ close to the horizon one obtains
    \begin{eqnarray}
x_\delta \approx x_H + \frac{\delta^2 x_C^2}{x_H (x_H - x_C)
(\varepsilon^2 x_H + x_C) }\,.
\label{xd}
\end{eqnarray}
    Note that smallness of the value of $x_\delta - x_H $ does not mean
the smallness of the ``physical distance'' from~$r_\delta$ to horizon
(see Ref.~\cite{LL_II}, Sec.~84).

\begin{figure}[h]
\centering
\includegraphics[height=40mm]{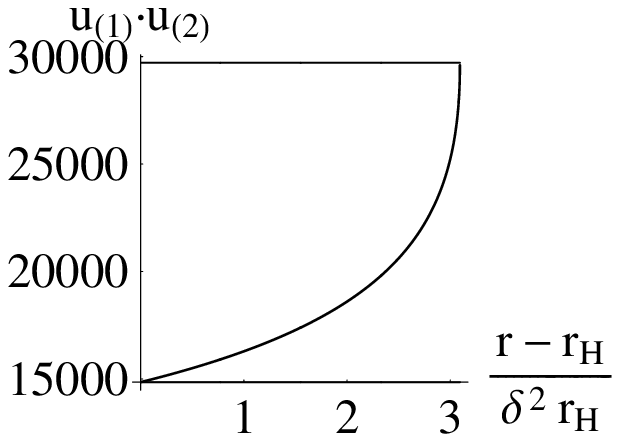}
\includegraphics[height=38mm]{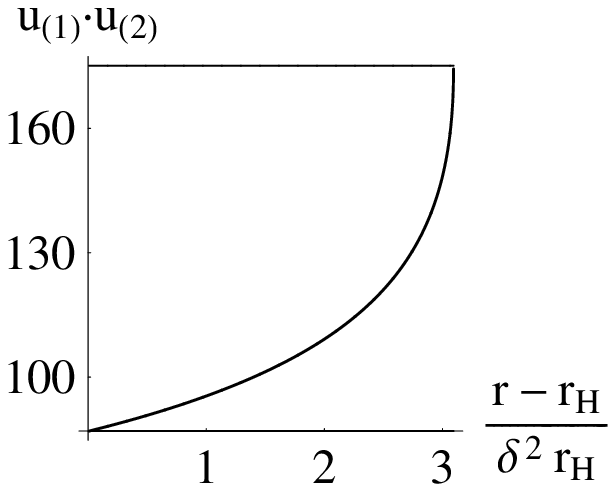}
\caption{The dependence of $u_{(1)}^i u_{(2) i}$ on the coordinate~$r$
for $A=0.998, \
\delta = 5 \cdot 10^{-4},\ \varepsilon_1 =\varepsilon_2=1$, \
$l_2 = l_L$ on the left and $l_2=l_R$ on the right}
\label{FigLR}
\end{figure}

    From~(\ref{KerrL1L2}), (\ref{xd}) one obtains for
values of function $u_{(1)}^i u_{(2) i}$ for $l_1 = l_{1H}-\delta$
after first scattering in points $x_\delta$ and on the horizon
         \begin{eqnarray}
\label{FxdH0}
u_{(1)}^i u_{(2) i} (x_\delta) \approx
\frac{(l_{2H} - l_2)(\varepsilon_1^2 x_H + x_C)}{ \delta \cdot x_C},
\\
u_{(1)}^i u_{(2) i} (x_H) \approx
\frac{(l_{2H} - l_2)(\varepsilon_1^2 x_H + x_C)}{ 2 \delta \cdot x_C}.
\label{FxdH}
\end{eqnarray}
    Therefore the function $u_{(1)}^i u_{(2) i}$ and hereby the energy of
collisions decrease near the horizon!
    The dependence of $u_{(1)}^i u_{(2) i}$ on the coordinate~$r$ is
shown on Fig.~\ref{FigLR}.

    The left picture shows that the collision energy can be very large
immediately after obtaining the angular momentum $ l_H - \delta$.
    But, for $A_{\rm max} =0.998 $, \ $ l_H - l_R \approx 0.04$ and so the
collision energy for particles with angular momentum $l_R$ is not large.
    This means that from the decrease of the collision energy with fixed
value of angular momentum does not follow the extremely large energy of
particles in the centre of mass frame needed to get the value $l_H - \delta$
in the intermediate collision.

    The decrease of the collision energy in the centre of mass frame of
the free falling particle and the particle with critical angular
momentum~$l=l_H-\delta$ in their movement to horizon is explained by
the decrease of the relative velocity of these particles when going from
the point~$x_\delta$.
    Due to the definition~(\ref{xd}) the radial velocity of the particle
with the angular momentum~$l=l_H-\delta$ is equal to zero:
$ d r / d \tau =0$ and $ d r / d t =0$.
     But the other noncritical particle colliding with the critical one
has large value of the radial velocity.
    Angular velocities $ d \varphi / d t $ for both particles go to
the same value $\Omega_H$.
    So large relative velocity of two particles occurs due to a ``stop''
of the critical particle in radial direction.
    But after this both particles increase their radial velocities
$ d r / d \tau $ and the relative velocity is decreasing --- the critical
particle is ``running down'' the free falling one.
    From~(\ref{Relsk02}), (\ref{FxdH0}), (\ref{FxdH}) one obtains for
the relative velocity of the particles colliding at the point~$x_\delta$
and on the horizon
             \begin{equation} \label{xdrel}
1 - v_{\rm rel}(x_\delta) =
\frac{\delta^2 x_C^2}{2 \left( l_{2H} - l_2 \right)^2\!
\left(\varepsilon_1^2 x_H + x_C \right)^2},
\end{equation}
             \begin{equation} \label{xHrel}
1 - v_{\rm rel} (x_H ) =
\frac{2 \delta^2 x_C^2}{\left( l_{2H} - l_2 \right)^2\!
\left(\varepsilon_1^2 x_H + x_C \right)^2}.
\end{equation}

    So the physical reason of the unlimited great energy of the collision
in the centre of mass frame of the particles falling in the black hole is
the increasing of the relative velocity of particles at the moment of
collision to the velocity of light.
    So one can expect very large energy of collision in the case when one
of the particles due to multiple intermediate collisions in the accretion
disc strongly diminishes its energy so that its velocity becomes small near
the horizon.
    Really from~(\ref{KerrLime1e2e2f}) it is easy to obtain that
    \begin{equation} \label{xHe1malo}
\frac{E_{\rm c.m.}}{\sqrt{m_1 m_2}} \sim
\sqrt{\frac{l_{2H} - l_2}{l_{1H} -l_1} } \to \infty\,,
\ \ \ \varepsilon_1 , l_1 \to 0 \,.
\end{equation}

    From the same considerations one can conclude that for the case of
nonrotating Schwarzschild black hole the collision energy of the free
falling particle with the particle at rest close to horizon also is
great and unlimited.
    Using~(\ref{SCM2af}), (\ref{KerrL1L2}), (\ref{bh1}) for $A=0$ and
the particle at rest in the point with radial coordinate~$r_0$
(so $l_1=0$, $\varepsilon_1 = \sqrt{1-r_g/r_0}$, $d r_1/d \tau=0$)
for the energy of its collision with the particle with $\varepsilon_2, l_2$
one obtains in the centre of mass frame
     \begin{equation} \label{ESchw}
E_{\rm c.m.}^{\,2} = m_1^2 + m_2^2 + 2 m_1 E_2
\sqrt{ \frac{r_0}{r_0- r_g}},
\end{equation}
    which evidently is growing infinitely for $r_0 \to r_g$.

    Note that the stopping of one particle on the horizon
of a nonrotating charged black hole takes place for its critical charge
and then it follows to expect the infinity energy of collisions as it was
shown in~\cite{Zaslavskii10b}.

    If one particle in the point with radial coordinate~$r_0$
has $d r /d \tau=0$ but $l_1 \ne 0$, then from
(\ref{SCM2af}), (\ref{KerrL1L2}), (\ref{geodKerr3dop}) one has
     \begin{eqnarray}
E_{\rm c.m.}^{\,2} = \hspace{77pt}
\nonumber \\
m_1^2 \!+ m_2^2 + 2 m_1 m_2 \left[ \varepsilon_2
\sqrt{ \frac{l_1^2 + x_0^2}{(x_0 \!-2) x_0}} -\frac{l_1 l_2}{x_0^2} \right]\!,\,
\label{ESchwl}
\end{eqnarray}
    which also is growing infinitely for $x_0 \to x_H=2$.

    Note that for particles nonrelativistic on infinity with $m_1=m_2=m$,
freely falling on the Schwarzschild black hole the limiting energy of
collisions is only $2 \sqrt{5} m$ (see Ref.~\cite{Baushev09}).

    In conclusion of this part note that
the probability of the collision of the relativistic particle
with the particle at rest close to the Schwarzschild horizon is very small.
    So this is the main difference with the situation when
the BSW resonance occurs.
    This can be seen from the evaluation of the interval of the proper time
of falling from the point~$r_0$ where $dr/ d \tau =0$ to horizon.
    Define the effective potential through the right hand side
of~(\ref{geodKerr3dop})
    \begin{equation} \label{Leff}
\frac{1}{2} \left( \frac{d r}{d \tau} \right)^2 + V_{\rm eff} = 0.
\end{equation}
    Then
    \begin{equation} \label{tau}
\frac{d r}{ d \tau}\Bigl|_{r_0} = 0 \ \ \Rightarrow \ \
r \approx
r_0 - \frac{\Delta \tau^2}{2} \frac{d V_{\rm eff}}{d r} \,.
\end{equation}
    So the proper time of falling of the particle to horizon is
    \begin{equation} \label{tau2}
\Delta \tau \approx M \sqrt{ 2(x_0 - x_H) \biggl/
\frac{d V_{\rm eff}}{d x}}\,.
\end{equation}
    For the Schwarzschild black hole one obtains
    \begin{equation} \label{tau3}
\Delta \tau \approx 4 M \sqrt{ \frac{2(x_0 - x_H)}
{4 + l^2} }\,, \ \ x_0 \to x_H .
\end{equation}
    For the Kerr black hole taking $l$ close to $l_H$
from (\ref{geodKerr3dop}), (\ref{Leff}), (\ref{tau2}) one obtains
    \begin{equation} \label{tau4}
\hspace*{-7pt}
\Delta \tau \approx M A \sqrt{ \frac{2 x_H (x_0 - x_H)}
{\sqrt{1\!-\! A^2} \left( \varepsilon^2 x_H \!+ x_C \right)}},
\ \ x_0 \! \to x_H ,\,
\end{equation}
    which evidently is much larger than~(\ref{tau3}) for $A\to 1$.

\section{Estimate of the time of the fall before collision leading to
the large energy}

    In Refs.~\cite{GribPavlov2010}--\cite{GribPavlov2010c}
it was shown by us that in order to get the unboundedly growing energy
for the extremal case one must
have the time interval (as coordinate as proper time) from the beginning
of the falling inside the black hole to the moment of collision also growing
infinitely.
    Quantitative estimations have been given for a case of extremely rotating
black hole $A=1$~\cite{GribPavlov2011,GribPavlov2011b}.
    Here we consider the case $A<1$.

    From Eqs.~(\ref{geodKerr1}), (\ref{geodKerr3dop}) one gets
    \begin{eqnarray}
\left| dt /dx \right| = \hspace{55pt}
\nonumber \\
\frac{ M \sqrt{x} \left((x^3 + A^2 x + 2 A^2) \varepsilon - 2 A l \right)}
{\Delta_x \sqrt{ 2 \varepsilon^{\mathstrut 2} x^2 \!- l^2 x +
2 (A \varepsilon \!- l)^2 + (\varepsilon^2 \!- 1) \Delta_x}} .
 \label{telHe}
\end{eqnarray}
    For $A<1$ from~(\ref{Delxxhxc}), (\ref{telHe}) one gets that
the value of the time interval measured by clock of the distant observer
necessary to achieve the horizon is logarithmically divergent.
    From~(\ref{telHe}) one has for $l< l_H=2 \varepsilon x_H/A $,\, $A<1$
and $x_0$ close to $x_H$ (for example $x_0=2 x_H$)
    \begin{equation} \label{telHelt}
\Delta t \sim - \frac{2 M x_H}{x_H - x_C} \log (x_f - x_H) \,, \ \ \
x_f \to x_H .
\end{equation}
    Remind that for the extremal black hole and the critical value of the
angular momentum of the falling particle this interval is divergent
as $ 1/(r-r_H)$.

    From Eqs.~(\ref{SCM2af}), (\ref{xd}), (\ref{FxdH0}), (\ref{FxdH}),
(\ref{telHelt}) it is easy to obtain for collisions of two particles with
$ l_1=l_H -\delta $ close to horizon at the point $x_\delta$
    \begin{equation} \label{tt3}
\Delta t \sim \frac{8 M x_H}{x_H-x_C} \, \log \frac{E_{\rm c.m.}}
{\sqrt{m_1 m_2}}\,.
\end{equation}
    So for
    \begin{equation} \label{tt4}
\hspace*{-7pt}
A=0.998, \ \
\Delta t \sim 3.2 \cdot 10^{-4} \frac{M}{M_\odot}
\log\! \frac{E_{\rm c.m.}}{\sqrt{m_1 m_2}} \, {\rm s}.\,
\end{equation}
    Taking the value of the Grand Unification energy
$E_{\rm c.m.}/\sqrt{m_1 m_2}=10^{14}$ and the mass of
the black hole $10^8 M_\odot$ typical for Active Nuclei of galaxies
one obtains $\Delta t \sim 10^6$\,s, i.e. of the order of 12 days.
    So in case of the nonextremal rotating black hole the mechanism of
the intermediate collision to get the additional angular momentum with
the following collision with other relativistic particle leading to large
collision energy proposed by us in Ref.~\cite{GribPavlov2010}
needs reasonable time much smaller than that for the extremal case.

    One can ask about the time of back movement of the particle after
collision with very high energy from the vicinity of horizon to the Earth.
    Due to reversibility of equations of motion in time it is easy to see
that this time is equal to the sum of the same 12 days to accretion disc
and some 10--100 megaparsec --- the distance of the AGN from the Earth.

{\bf Acknowledgments.}
    One of the authors A.A.G. is indebted to CAPES for financial support of
the work on the first part of the paper and to the UFES, Brazil,
for hospitality, other  author O.F.P. is thankful to CNPq for
partial financial support of his work.


\end{document}